\documentclass[onside,12pt,onecolumn,conference]{IEEEtran}

\usepackage[utf8]{inputenc}
\usepackage[T1]{fontenc}

\usepackage{amsmath}
\usepackage{amssymb}
\usepackage{url}
\usepackage{xspace}
\usepackage{color}
\usepackage{listings}
\usepackage{tikz}
\usepackage{url}
\usepackage{booktabs} 
\usepackage{subcaption}
\usepackage{enumitem}
\usepackage{microtype}
\usepackage{ifthen}
\usepackage{caption}
\usepackage{hyperref}

\usepackage[numbers]{natbib}
\usepackage[binary-units]{siunitx}

\newcommand{\List}{Listing\xspace}
\newcommand{\fig}{Figure\xspace}

\newcommand{\ie}{i.e.\xspace}
\newcommand{\eg}{e.g.\xspace}

\newcommand{\schedjob}{\texttt{Job}\xspace}
\newcommand{\schedmachine}{\texttt{Machine}\xspace}
\newcommand{\schedjobassignment}{\texttt{JobAssignment}\xspace}
\newcommand{\schedschedule}{\texttt{Schedule}\xspace}

\lstdefinelanguage{Julia}%
  {morekeywords={abstract,break,case,catch,const,continue,do,else,elseif,%
      end,export,false,for,function,immutable,import,importall,if,in,%
      macro,module,otherwise,quote,return,switch,true,try,type,typealias,%
      using,while},%
   sensitive=true,%
   alsoother={$},
   morecomment=[l]\#,%
   morecomment=[n]{\#=}{=\#},%
   morestring=[s]{"}{"},%
   morestring=[m]{'}{'},%
}[keywords,comments,strings]%

\newcommand{\Icmax}{\ensuremath{C_{\text{max}}}\xspace}
\newcommand{\Igraham}[3]{\ifthenelse{\equal{#2}{}}{#1\,$|$$|$\,#3}{#1\,$|$\,#2\,$|$\,#3}}

\newcommand{\schedjl}{\texttt{Scheduling.jl}\xspace}

\author{\IEEEauthorblockN{Sascha Hunold}
\IEEEauthorblockA{TU Wien\\
Faculty of Informatics\\
Vienna, Austria\\
Email: hunold@par.tuwien.ac.at}
\and
\IEEEauthorblockN{Bartłomiej Przybylski}
\IEEEauthorblockA{Adam Mickiewicz University\\
Faculty of Mathematics and Computer Science\\
Poznań, Poland\\
Email: bap@amu.edu.pl}
}

\begin{document}

\title{Scheduling.jl -- Collaborative and Reproducible Scheduling Research with Julia}

\maketitle

\begin{abstract}
  We introduce the \schedjl{} Julia package, which is intended for
  collaboratively conducting scheduling research and for sharing
  implementations of algorithms. It provides the fundamental building
  blocks for implementing scheduling algorithms following the
  three-field notation of Graham et al., i.e., it has functionality to
  describe machine environments, job characteristics, and optimality
  criteria.  Our goal is to foster algorithm and code sharing in the
  scheduling community. \schedjl can also be used to support teaching
  scheduling theory in classes. We will show the main functionalities
  of \schedjl and give an example on how to use it by comparing
  different algorithms for the problem of \Igraham{P}{}{\Icmax}.
\end{abstract}


\section{Introduction}

Scheduling is one of the fundamental research subjects, which is
central to virtually all scientific domains that require any kind of
resource sharing. Therefore, a large body of literature exists that
introduces basic scheduling algorithms for various scheduling
problems~\cite{Blazewicz2019,brucker2004,Drozdowski09,handbook2004,pinedo2016}.
Most theoretical works in scheduling research use the three-field
notation $\alpha|\beta|\gamma$ of \citet{graham1979optimization} for
classifying scheduling problems.  By using this notation, each
scheduling problem can be described by the machine environment
$\alpha$, the job characteristics $\beta$, and the optimality
criterion $\gamma$. For example, \Igraham{P}{}{\Icmax} defines the
problem of scheduling jobs on identical parallel machines, where the
maximum completion time (\Icmax) should be minimized, while no further
job characteristics are given. One example of such job characteristics
could be the \emph{moldable} job model (denoted as $any$), \ie, in the
problem
\Igraham{P}{any}{\Icmax}~\cite{Drozdowski09,DBLP:journals/concurrency/Hunold15}. In
the moldable model, each job can not only be executed on \num{1}
machine, but it may be allotted to several machines (between \num{1}
and $m$ machines). The number of machines is selected by the
scheduler, but this number of machines will not change until a job has
been completed. Another example of job characteristics are job
processing times that are variable and depend on environmental factors
such as the position $r$ of a job in a schedule. For example, in the
\Igraham{P}{in-tree, $p_{j,r} = \varphi(r)$}{\Icmax} problem, the
processing times of jobs with in-tree precedence constraints are
described by the $\varphi$ function
\cite{Przybylski2017,Przybylski2018a}.

Since scheduling problems are so fundamental to many scientific
disciplines, thousands of algorithms exist for a seemingly endless
list of problem variations. Among them, a significant number of
scheduling algorithms can be described in the three-field notation.
Several surveys on specific scheduling problems, \eg,
\Igraham{P}{}{\Icmax}, have been conducted that compare the scheduling
performance of different algorithms via simulations. Although these
studies are very informative for the readers, they are often hard to
reproduce, as many of the building blocks are imprecisely explained
and because the source code is often not provided or has become
inaccessible over the years. For that reason, many algorithms cannot
be compared fairly or in a scientifically sound manner, as too many
details are missing.

To overcome this problem, we propose \schedjl, which provides a
generic and open scheduling platform, on top of which a large number
of scheduling algorithms can be implemented.

In the remainder of the article, we introduce the core functionalities
of the \schedjl package and show an example of how to use them.

\section{Overview of \schedjl}

\schedjl provides the main building blocks for implementing scheduling
algorithms in their most generic form, which are \schedjob,
\schedmachine, \schedjobassignment, and \schedschedule.  A classical
\schedjob $J_j$ is defined by its processing time $p_j$ but can also be
characterized by a weight $w_j$, a release date $r_j$, a due date
$d_j$, or a deadline $\bar{d_j}$. A \schedmachine $M_i$ is mainly
defined by its speed. Of course, the sets of parameters used can be easily extended. The task of a scheduling algorithm is to find an
assignment of jobs to machines, such that a given criterion is
optimized. An assignment of jobs to machines defines the starting and
the completion time of a job $J_j$ on a machine $M_i$. The final
\schedschedule is composed of a vector of jobs, a vector of machines,
and a vector of job to machine assignments. It is worth noticing that \schedjl is designed to operate on exact values (rational numbers) rather than inexact ones (floating point numbers).

Once a schedule has been obtained by executing an algorithm, the
package \schedjl provides different optimization criteria that can be
computed for a schedule, e.g., the makespan \Icmax, the average
completion time $\sum_j C_j$, or the number of tardy jobs
$\sum_j U_j$.  Additionally, the package is shipped with
implementations of various scheduling algorithms, in particular, for
\Igraham{P}{}{\Icmax}.

In order to obtain \schedjl, one needs to install the package from
\url{https://github.com/bprzybylski/Scheduling.jl}.  The stable
version can be installed by calling \verb|Pkg.add("Scheduling")| and
the development version by executing \verb|Pkg.develop("Scheduling")|.

\List~\ref{lst:example} presents an example of how to leverage the
basic functionality of \schedjl. Here, we create a set of jobs $J$ and
a set of machines $M$. Then, we can apply the LPT algorithm to obtain
a schedule.  On this schedule, we can compute various metrics like
\Icmax or $\sum_j C_j$.

\begin{lstlisting}[float=t,caption={Example of using the basic functionality of \schedjl; applying the LPT algorithm to a small scheduling problem and reporting the \Icmax and the $\sum_j C_j$ metrics.},label={lst:example}]
using Scheduling
using Scheduling.Algorithms
using Scheduling.Objectives

# Generate a set of jobs with processing times
J = Jobs([27, 19, 19, 4, 48, 38, 29])
# Generate a set of 4 identical machines
M = Machines(4)
# Generate a schedule using LPT list rule
LPT = Algorithms.lpt(J, M)
println("Cmax     = $(Int(cmax(LPT)))")
println("sum(C_j) = $(Int(csum(LPT)))")
\end{lstlisting}

\schedjl also provides means to visualize the resulting schedules. For
example, scientists can choose to produce an image of a schedule,
where it is also possible to animate the schedule creation. If
desired, the schedule can also be plotted as an TikZ image, which can
directly be inserted into publications.

\section{Using \schedjl: The Case of \texttt{$P||\Icmax$}}

Now, we turn our attention to the NP-hard problem
\Igraham{P}{}{\Icmax}, for which \citet{DBLP:conf/afips/Graham72}
devised two fundamental approximation algorithms, namely the LIST and
the LPT (Largest Processing Time) algorithm.
\citet{DBLP:conf/afips/Graham72} showed that for any instance of
\Igraham{P}{}{\Icmax}, LIST provides a $2-1/m$ approximation, while
LPT improves this bound to
$4/3-1/(3m)$. \citet{DBLP:journals/jacm/HochbaumS87} devised a PTAS
for this problem, by developing a dual approximation algorithm to
solve \Igraham{P}{}{\Icmax}, which internally relies on solving a
bin packing problem.

Although our list is far from exhaustive, we discuss several
heuristics that have been proposed to solve \Igraham{P}{}{\Icmax}.
\citet{DBLP:journals/cor/FrancaGLM94} developed the 3-PHASE algorithm
for which the authors stated that it ``outperforms alternative
heuristics'' on their respective test instances. Similarly,
\citet{DellAmico08} presented heuristics that are combined to compute
exact solutions for various instances of \Igraham{P}{}{\Icmax}.
Last, \citet{Ghalami19} presented a parallel implementation of the
algorithm of \citet{DBLP:journals/jacm/HochbaumS87}. In each of these
works, the authors implemented their own interpretation of existing
algorithms.  They also generated their own problem instances and
reported results for a subset of the instances. Neither the
implementations of the algorithms nor the instances (or generators)
are available anymore.  Such a lack of code and meta-information is a
common problem in many scientific fields when looking at the
reproducibility of results. Independent researchers, who would like to
continue studying heuristics for \Igraham{P}{}{\Icmax}, will have to
start from scratch and create implementations and instances
themselves.

We have developed \schedjl to improve the reproducibility in the
scheduling domain. For the problem of \Igraham{P}{}{\Icmax},
\schedjl contains several algorithms that can be used to solve given
instances.  Besides heuristics with guarantees (\ie, LIST and LPT),
\schedjl also contains an implementation of the algorithm of
\citet{DBLP:journals/jacm/HochbaumS87} as well as an exact algorithm
that was presented by \citet{Drozdowski09}.

\begin{figure*}[t]
  \centering
  \begin{subfigure}[t]{.7\linewidth}
    \centering
    \includegraphics[width=\linewidth]{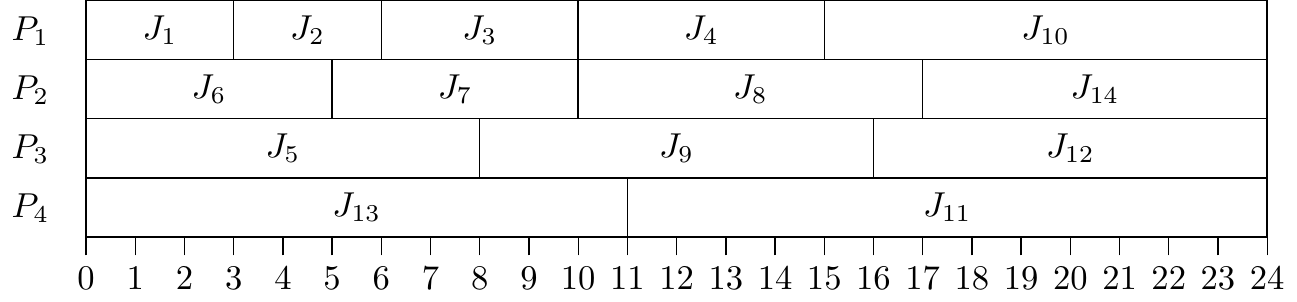}
    \subcaption{\label{fig:alg_hs}Hochbaum\,\&\,Shmoys algorithm}
  \end{subfigure}\\[1ex]
  \begin{subfigure}[t]{.7\linewidth}
    \centering
    \includegraphics[width=\linewidth]{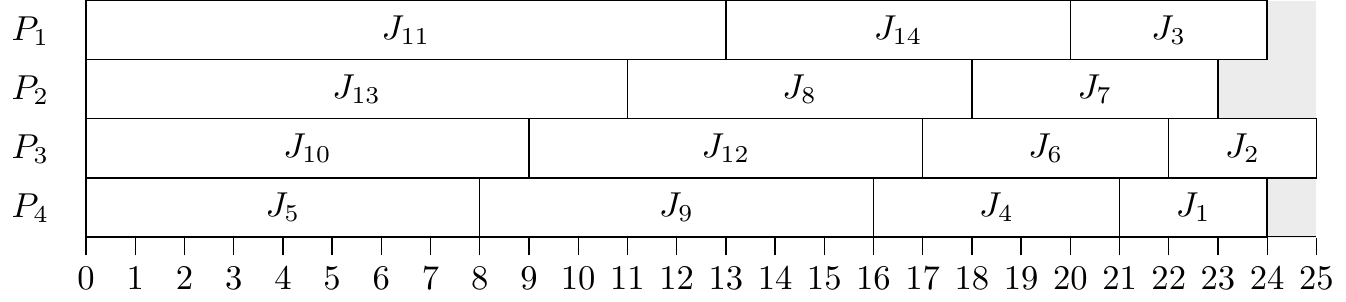}
    \subcaption{\label{fig:alg_lpt}LPT algorithm}
  \end{subfigure}\\[1ex]
  \begin{subfigure}[t]{.7\linewidth}
    \centering
    \includegraphics[width=\linewidth]{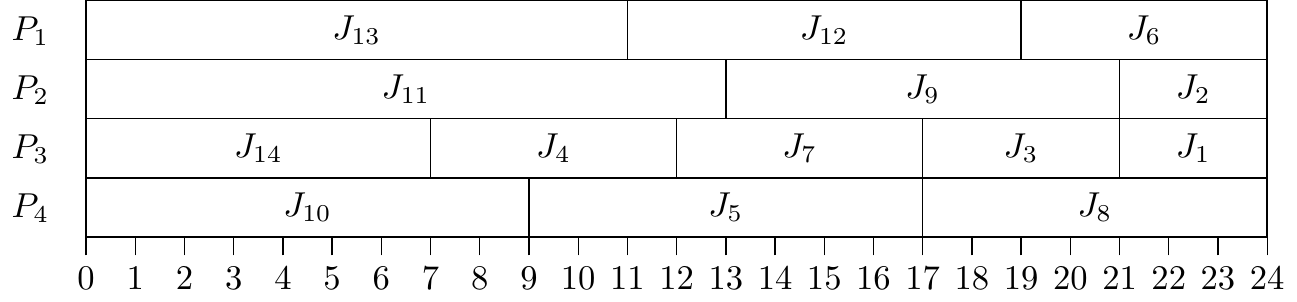}
    \subcaption{\label{fig:alg_opt}OPT (exact)}
  \end{subfigure}
  \caption{\label{fig:pcmax_schedules}Comparing schedules produced by
    different algorithms for P$||\Icmax$, where Gantt charts of
    schedules have been created with \schedjl.}
\end{figure*}

If independent researchers now set out to compare novel heuristics to
already established methods, they could simply use the source code
provided. \List~\ref{lst:schedules} exemplifies how different
algorithms for one specific instance of a problem can be compared.  In
this particular case, we created three different TiKZ files containing
the Gantt charts of the schedules. \fig~\ref{fig:pcmax_schedules}
presents these Gantt charts produced by different algorithms when
solving an instance of \Igraham{P}{}{\Icmax}.

\begin{lstlisting}[float=t,caption={Comparing different algorithms for an instance of P$||\Icmax$.},label={lst:schedules}]
J = Jobs([3, 3, 4, 5, 8, 5, 5, 7, 8, 9, 13, 8, 11, 7])
M = Machines(4)

S_LPT = Algorithms.lpt(J, M)
S_OPT = Algorithms.P__Cmax_IP(J, M)
S_HS  = Algorithms.P__Cmax_HS(J, M; eps=1//10)
Scheduling.TeX(S_LPT, "schedule_lpt.tex")
Scheduling.TeX(S_OPT, "schedule_opt.tex")
Scheduling.TeX(S_HS, "schedule_hs.tex")
\end{lstlisting}

\section{Conclusions and Future Work}

We have introduced the Julia package \schedjl, which is an effort to
increase the reproducibility in the scheduling community. By providing
basic building blocks for developing scheduling methods as well as
several implementations of well-known scheduling algorithms, \schedjl
can serve as a foundation for developing a large variation of
scheduling algorithms. The package provides easy-to-use plotting
functions to easily obtain Gantt charts of computed schedules.

The package is far from complete and it should serve as a starting
point for future work. So far, we have focused on providing a general
development platform for implementing various algorithms. We made sure
that design choice are applicable by implementing multiple algorithms
for the classical problem of \Igraham{P}{}{\Icmax} ourselves.  Now,
we hope that the community will contribute new algorithms to this
package. We will also continue to integrating more algorithms into
\schedjl, starting with algorithms for which Julia code already exists,
e.g., the algorithm of \citet{DBLP:journals/tpds/BleuseHKMMT17} for
\Igraham{(P$m$,P$k$)}{mold}{$C_{\max}$}.

\bibliographystyle{plainnat}
\bibliography{main}
\end{document}